\documentclass[twocolumn,pre,superscriptaddress]{revtex4}

\usepackage{graphicx}
\usepackage{amsmath}
\usepackage{subfigure}
\usepackage{calrsfs}

\newcommand\mat{\mathbf}

\newcommand\half{\mbox{$\frac12$}}

\newcommand{\e}{\mathrm{e}}

\begin{document}

\title{Friendship networks and social status}

\author{Brian Ball}
\affiliation{Department of Physics, University of Michigan, Ann Arbor,
MI 48109, U.S.A.}
\author{M. E. J. Newman}
\affiliation{Department of Physics, University of Michigan, Ann Arbor,
MI 48109, U.S.A.}
\affiliation{Center for the Study of Complex Systems, University of
  Michigan, Ann Arbor, MI 48109, U.S.A.}

\begin{abstract}
  In empirical studies of friendship networks participants are typically
  asked, in interviews or questionnaires, to identify some or all of their
  close friends, resulting in a directed network in which friendships can,
  and often do, run in only one direction between a pair of individuals.
  Here we analyze a large collection of such networks representing
  friendships among students at US high and junior-high schools and show
  that the pattern of unreciprocated friendships is far from random.  In
  every network, without exception, we find that there exists a ranking of
  participants, from low to high, such that almost all unreciprocated
  friendships consist of a lower-ranked individual claiming friendship with
  a higher-ranked one.  We present a maximum-likelihood method for deducing
  such rankings from observed network data and conjecture that the rankings
  produced reflect a measure of social status.  We note in particular that
  reciprocated and unreciprocated friendships obey different statistics,
  suggesting different formation processes, and that rankings are
  correlated with other characteristics of the participants that are
  traditionally associated with status, such as age and overall popularity
  as measured by total number of friends.
\end{abstract}

\maketitle

\subsection*{Introduction}

A social network, in the most general sense of the term, consists of a
group of people, variously referred to as nodes or actors, connected by
social interactions or ties of some kind~\cite{WF94}.  In this paper we
consider networks in which the ties represent friendship.  Friendship
networks have been the subject of scientific study since at least the
1930s.  A classic example can be found in the studies by Rapoport and
collaborators of friendship among schoolchildren in the town of Ann Arbor,
MI in the 1950s and 60s~\cite{RH61}, in which the investigators circulated
questionnaires among the students in a school asking them to name their
friends.  Many similar studies have been done since then, with varying
degrees of sophistication, but most employ a similar questionnaire-based
methodology.  A counterintuitive aspect of the resulting networks is that
they are directed.  Person~A states that person~B is their friend and hence
there is a direction to the ties between individuals.  It may also be that
person~B states that person~A is their friend, but it does not have to be
the case, and in practice it turns out that a remarkably high fraction of
claimed friendships are not reciprocated.  In the networks we study in this
paper the fraction of reciprocated ties rarely exceeds 50\% and can be as
low as 30\%.

This could be seen as a problem for the experimenter.  One thinks of
friendship as a two-way street---a friendship that goes in only one
direction is no friendship at all.  How then are we to interpret the many
unreciprocated connections in these networks?  Are the individuals in
question friends or are they not?  One common approach is simply to
disregard the directions altogether and consider two individuals to be
friends if they are connected in either direction (or
both)~\cite{Airoldi11}.  In this paper, however, we take a different view
and consider what we can learn from the unreciprocated connections.  It has
been conjectured that, rather than being an error or an annoyance, the
pattern of connections might reflect underlying features in the structure
or dynamics of the community under study~\cite{Homans50,Davis72,DBF00}.

Working with a large collection of friendship networks from US schools, we
find that in every network there is a clear ranking of individuals from low
to high such that almost all friendships that run in only one direction
consist of a lower-ranked individual claiming friendship with a
higher-ranked one.  We conjecture that these rankings reflect a measure of
social status and present a number of results in support of this idea.  For
instance, we find that a large majority of reciprocated friendships are
between individuals of closely similar rank, while a significant fraction
of unreciprocated friendships are between very different ranks, an
observation consistent with qualitative results in the sociological
literature going back several decades~\cite{Davis72}.  We also investigate
correlations between rank and other individual characteristics, finding,
for example, that there is a strong positive correlation between rank and
age, older students having higher rank on average, and between rank and
overall popularity, as measured by total number of friends.

The outline of the paper is as follows.  First, we describe our method of
analysis, which uses a maximum-likelihood technique in combination with an
expectation--maximization algorithm to extract rankings from directed
network data.  Then we apply this method to school friendship networks,
revealing a surprisingly universal pattern of connections between
individuals in different schools.  We also present results showing how rank
correlates with other measures.  Finally we give our conclusions and
discuss possible avenues for future research.

\subsection*{Inference of rank from network structure}
\label{Sec:Math}
Consider a directed network of friendships between $n$ individuals in which
a connection running from person~A to person~B indicates that A claims B as
a friend.  Suppose that, while some of the friendships in the network may
be reciprocated or bidirectional, a significant fraction are
unreciprocated, running in one direction only, and suppose we believe there
to be a ranking of the individuals implied by the pattern of the
unreciprocated friendships so that most such friendships run from lower to
higher rank.  One possible way to infer that ranking would be simply to
ignore any reciprocated friendships and then construct a minimum violations
ranking of the remaining network~\cite{Reinelt85,ACK86}.  That is, we find
the ranking of the network nodes that minimizes the number of connections
running from higher ranked nodes to lower ranked ones.  In practice this
approach works quite well: for the networks studied in this paper the
minimum violations rankings have an average of~98\% of their unreciprocated
friendships running from lower to higher ranks and only~2\% running the
other way.  By contrast, versions of the same networks in which edge
directions have been randomized typically have about~10\% of edges running
the wrong way.  (Statistical errors in either case are 1\% or less, so
these observations are highly unlikely to be the results of chance.)

The minimum violations ranking, however, misses important network features
because it focuses only on unreciprocated friendships.  In most cases there
are a substantial number of reciprocated friendships as well, as many as a
half of the total, and they contain significant information about network
structure and ranking.  For example, as we will see, pairs of individuals
who report a reciprocated friendship are almost always closely similar in
rank.  To make use of this information we need a more flexible and general
method for associating rankings with network structure.  In this paper we
use a maximum likelihood approach defined as follows.

Mathematically we represent the distinction between reciprocated and
unreciprocated friendships in the network using two separate matrices.  The
symmetric matrix~$\mat{S}$ will represent the reciprocated
connections---undirected edges in graph theory terms---such that
$S_{ij}=S_{ji}=1$ if there are connections both ways between nodes $i$
and~$j$, and zero otherwise.  The asymmetric matrix~$\mat{T}$ will
represent the unreciprocated (directed) edges with $T_{ij}=1$ if there is a
connection to node~$i$ from node~$j$ (but not \textit{vice versa}), and
zero otherwise.  The matrices $\mat{S}$ and $\mat{T}$ are related to the
conventional adjacency matrix~$\mat{A}$ of the network by
$\mat{A}=\mat{S}+\mat{T}$.

Now suppose that there exists some ranking of the individuals, from low to
high, which we will represent by giving each individual a unique integer
rank in the range 1 to~$n$.  We will denote the rank of node~$i$ by $r_i$
and the complete set of ranks by~$R$.  We have found it to be a good
approximation to assume that the probability of friendship between two
individuals is a function only of the difference between their ranks.  We
specifically allow the probability to be different for reciprocated and
unreciprocated friendships, which acknowledges the possibility that the two
may represent different types of relationships, as conjectured for instance
in~\cite{Davis72,Dijkstra10}.  We define a function $\alpha(r_i-r_j)$ to
represent the probability of an undirected edge between $i$ and~$j$ and
another $\beta(r_i-r_j)$ for a directed edge to~$i$ from~$j$.  Since
$\alpha(r)$ describes undirected edges it must be symmetric
$\alpha(-r)=\alpha(r)$, but $\beta(r)$ need not be symmetric.

If we were not given a network but we were given the probability functions
$\alpha$ and $\beta$ and a complete set of rankings on $n$ vertices, then
we could use this model to generate---for instance on a computer---a
hypothetical but plausible network in which edges appeared with the
appropriate probabilities.  In effect, we have a random graph model that
incorporates rankings.  In this paper, however, we want to perform the
reverse operation: given a network we want to deduce the rankings of the
nodes and the values of the functions $\alpha$ and~$\beta$.  To put that
another way, if we are given a network and we assume that it is generated
from our model, what values of the rankings and probability functions are
most likely to have generated the network we observe?

This question leads us to a maximum likelihood formulation of our problem,
which we treat using an expectation--maximization (EM) approach in which
the ranks~$R$ are considered hidden variables to be determined and the
functions~$\alpha$ and~$\beta$ are parameters of the model.  Using a
Poisson formulation of the random network generation process, we can write
the probability of generation of a network~$G$ with rankings~$R$, given the
functions $\alpha$ and~$\beta$, as
\begin{align}
P(G,R|\alpha,\beta)
  &= \prod_{i>j}
     {[\alpha(r_i-r_j)]^{S_{ij}}\over S_{ij}!}\,\e^{-\alpha(r_i-r_j)}
     \nonumber\\
  &  \quad{}\times \prod_{i\ne j} {[\beta(r_i-r_j)]^{T_{ij}}\over T_{ij}!}\,
     \e^{-\beta(r_i-r_j)}.
\label{eq:likelihood}
\end{align}
Note that we have excluded self-edges here, since individuals cannot name
themselves as friends.  We have also assumed that the prior probability
of~$R$ is uniform over all sets of rankings.

The most likely values of the parameter functions~$\alpha$ and~$\beta$ are
now given by maximizing the marginal likelihood $P(G|\alpha,\beta) = \sum_R
P(G,R|\alpha,\beta)$, or equivalently maximizing its logarithm, which is
more convenient.  The logarithm satisfies the Jensen inequality
\begin{equation}
\log \sum_R P(G,R|\alpha,\beta)
  \ge \sum_R q(R) \log {P(G,R|\alpha,\beta)\over q(R)},
\label{eq:jensen}
\end{equation}
for any set of probabilities~$q(R)$ such that $\sum_R q(R) = 1$, with
the equality being recovered when
\begin{equation}
q(R) = {P(G,R|\alpha,\beta)\over \sum_R P(G,R|\alpha,\beta)}.
\label{eq:estep}
\end{equation}
This implies that the maximization of the log-likelihood on the left side
of~\eqref{eq:jensen} is equivalent to the double maximization of the right
side, first with respect to~$q(R)$, which makes the right side equal to the
left, and then with respect to $\alpha$ and~$\beta$, which gives us the
answer we are looking for.  It may appear that expressing the problem as a
double maximization in this way, rather than as the original single one,
makes it harder, but in fact that's not the case.

The right-hand side of~\eqref{eq:jensen} can be written as $\sum_R q(R)
\log P(G,R|\alpha,\beta) - \sum_R q(R) \log q(R)$, but the second term does
not depend on $\alpha$ or~$\beta$, so as far as $\alpha$ and~$\beta$ are
concerned we need consider only the first term, which is simply the
average~$\overline{\mathcal{L}}$ of the log-likelihood over the
distribution~$q(R)$:
\begin{equation}
\overline{\mathcal{L}} = \sum_R q(R) \log P(G,R|\alpha,\beta).
\end{equation}
Making use of Eq.~\eqref{eq:likelihood} and neglecting an
unimportant overall constant, we then have
\begin{align}
\overline{\mathcal{L}}
  &= \sum_R q(R) \sum_{i\ne j} \Bigl[ \half S_{ij} \log \alpha(r_i-r_j)
     + T_{ij} \log \beta(r_i-r_j) \nonumber\\
  &\hspace{8em}{} - \half\alpha(r_i-r_j) - \beta(r_i-r_j) \Bigr],
\label{eq:mstep1}
\end{align}
where we have used the fact that $\alpha(r)$ is a symmetric function.

This expression can be simplified further.  The first term in the sum is
\begin{align}
& \half \sum_R q(R) \sum_{i\ne j} S_{ij} \log \alpha(r_i-r_j) \nonumber\\
  &\hspace{3em}{} = \half\sum_z \sum_{i\ne j} S_{ij} q(r_i-r_j=z)
                    \log \alpha(z),
\label{eq:msimplify1}
\end{align}
where $q(r_i-r_j=z)$ means the probability within the distribution~$q(R)$
that $r_i-r_j=z$.  We can define
\begin{equation}
a(z) = {1\over n-|z|} \sum_{i\ne j} S_{ij} q(r_i-r_j=z),
\label{eq:defsnu}
\end{equation}
which is the expected number of undirected edges in the observed network
between pairs of nodes with rank difference~$z$.  It is the direct
equivalent in the observed network of the quantity~$\alpha(z)$, which is
the expected number of edges in the model.  The quantity~$a(z)$,
like~$\alpha(z)$, is necessarily symmetric, $a(z)=a(-z)$, and
hence~\eqref{eq:msimplify1} can be written as
\begin{equation}
\half \sum_R q(R) \sum_{i\ne j} S_{ij} \log \alpha(r_i-r_j)
  = \sum_{z=1}^{n-1} (n-z) a(z) \log\alpha(z).
\end{equation}
Similarly, we can define
\begin{equation}
b(z) = {1\over n-|z|} \sum_{i\ne j} T_{ij} q(r_i-r_j=z)
\label{eq:defsnd}
\end{equation}
and
\begin{align}
&\sum_R q(R) \sum_{i\ne j} T_{ij} \log \beta(r_i-r_j) \nonumber\\
&\hspace{2em}{} = \sum_{z=1}^{n-1} (n-z)
                  \bigl[ b(z) \log\beta(z) + b(-z) \log\beta(-z) \bigr],
\end{align}
where $b(z)$ is the expected number of directed edges between a pair of
nodes with rank difference~$z$.  Our final expression for
$\overline{\mathcal{L}}$ is
\begin{align}
\overline{\mathcal{L}} &= \sum_{z=1}^{n-1} (n-z) \bigl[ a(z) \log\alpha(z)
  - \alpha(z) \nonumber\\
  &\quad{} + b(z) \log\beta(z) - \beta(z)
   + b(-z) \log\beta(-z) - \beta(-z) \bigr].
\end{align}
Our approach involves maximizing this expression with respect to
$\alpha(z)$ and $\beta(z)$ for given $a(z)$ and $b(z)$, which can be done
using standard numerical methods.  (Note that the expression separates into
terms for the directed and undirected edges, so the two can be maximized
independently.)  The values of $a(z)$ and $b(z)$ in turn are calculated
from Eqs.~\eqref{eq:estep}, \eqref{eq:defsnu}, and~\eqref{eq:defsnd},
leading to an iterative method in which we first guess values for
$\alpha(z)$ and~$\beta(z)$, use them to calculate~$q(R)$ and hence $a(z)$
and~$b(z)$, then maximize~$\overline{\mathcal{L}}$ to derive new values
of~$\alpha$ and~$\beta$, and repeat to convergence.  This is the classic
expectation--maximization approach to model fitting.

Two further elements are needed to put this scheme into practice.  First,
we need to specify a parametrization for the functions~$\alpha$
and~$\beta$.  We have found the results to be robust to the choice of
parametrization, but in the results reported here we find $\alpha$ to be
well represented by a Gaussian centered at the origin.  The
function~$\beta$ takes a more complicated form which we parametrize as a
Fourier cosine series, keeping five terms and squaring to enforce
nonnegativity, plus an additional Gaussian peak at the origin.

Second, the sum in the denominator of Eq.~\eqref{eq:estep} is too large to
be numerically tractable, so we approximate it using a Markov chain Monte
Carlo method---we generate complete rankings~$R$ in proportion to the
probability~$q(R)$ given by Eq.~\eqref{eq:estep} and average over them to
calculate~$a(z)$ and $b(z)$.

\subsection*{Results}
\label{Sec:Results}

We have applied the method of the previous section to the analysis of data
from the US National Longitudinal Study of Adolescent Health (the
``AddHealth'' study), a large-scale multi-year study of social conditions
for school students and young adults in the United States~\cite{note1}.
Using results from surveys conducted in 1994 and 1995, the study compiled
friendship networks for over $90\,000$ students in schools covering US
school grades 7 to 12 (ages 12 to 18 years).  Schools were chosen to
represent a broad range of socioeconomic conditions.  High schools (grades
9 to 12) were paired with ``feeder'' middle schools (grades 7 and 8) so
that networks spanning schools could be constructed.

To create the networks, each student was asked to select, from a list of
students attending the same middle/high school combination, up to ten
people with whom they were friends, with a maximum of five being male and
five female.  From these selections, 84 friendship networks were
constructed ranging in size from tens to thousands of students, one for
each middle/high school pair, along with accompanying data on the
participants, including school grade, sex, and ethnicity.  Some of the
networks divide into more than one strongly connected component, in which
case we restrict our analysis to the largest component only.  We perform
the EM analysis of the previous section on each network separately,
repeating the iterative procedure until the rankings no longer change.

\begin{figure}
\begin{center}
\includegraphics[width=\columnwidth]{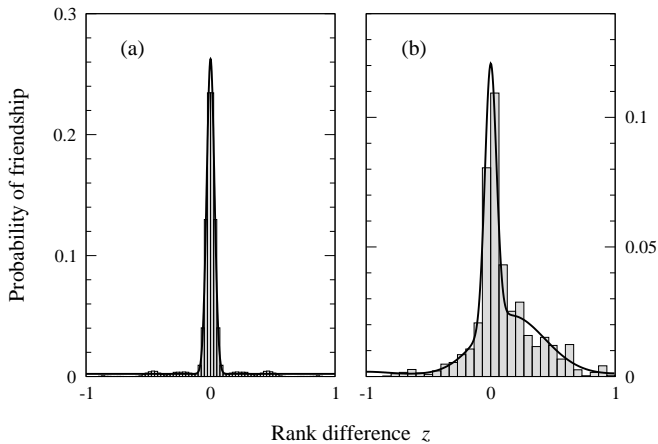}
\end{center}
\caption{(a)~Probability of reciprocated friendships as a function of rank
  difference (normalized to run from $-1$ to~1).  The histogram shows
  empirical results for a single example network; the solid curve is the
  fitted function~$\alpha(z)$.  (b)~The equivalent plot for unreciprocated
  friendships.}
\label{plot:p_data_show}
\end{figure}

Figure~\ref{plot:p_data_show} shows results for a typical network.  In
panel~(a), the histogram shows the measured value of the quantity~$a(z)$,
Eq.~\eqref{eq:defsnu}, the empirical probability of a reciprocated
friendship (technically the expected number of undirected edges) between a
vertex pair with rank difference~$z$, with the horizontal axis rescaled to
run from $-1$ to~$1$ (rather than $-n$ to~$n$).  As the figure shows the
probability is significantly different from zero only for small values
of~$z$, with a strong peak centered on the origin.  The solid curve shows
the fit of this peak by the Gaussian function~$\alpha(z)$, which appears
good.  The fit is similarly good for most networks.  The form of $a(z)$
tells us that most reciprocated friendships fall between individuals of
closely similar rank: there is a good chance that two people with roughly
equal rank will both claim the other as a friend, but very little chance
that two people with very different ranks will do so.  This result seems at
first surprising, implying as it does that people must be able to determine
their own and others' rank with high accuracy in order to form friendships,
but a number of previous studies have suggested that indeed this is
true~\cite{Anderson06}.

Panel~(b) of Fig.~\ref{plot:p_data_show} shows~$b(z)$,
Eq.~\eqref{eq:defsnd}, for the same network, which is the probability of a
directed edge between nodes with rank difference~$z$.  Again there is a
strong central peak to the distribution, of width similar to that for the
undirected edges, indicating that many unreciprocated friendships are
between individuals of closely similar rank.  However, the distribution
also has a substantial asymmetric tail for positive values of the rank
difference, indicating that in a significant fraction of cases individuals
claim friendship with those ranked higher than themselves, but that those
claims are not reciprocated.  The black curve in the panel shows the best
fit to the function~$\beta(z)$ in the maximum-likelihood calculation.

\begin{figure}
\begin{center}
\includegraphics[width=0.98\columnwidth]{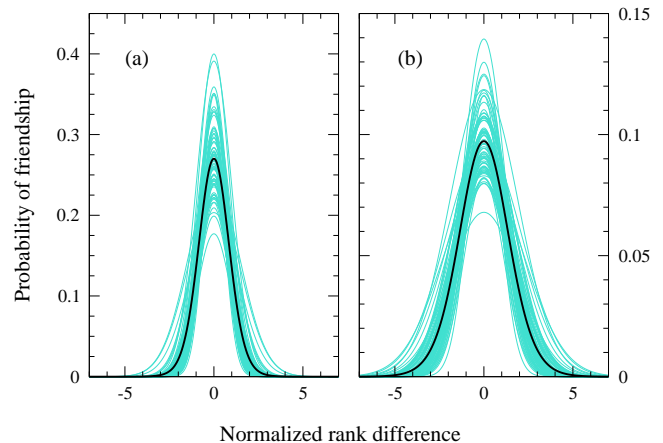}
\end{center}
\caption{The fitted central peak of the friendship probability
  distributions for (a)~reciprocated and (b)~unreciprocated friendships.
  The horizontal axes are measured in units of absolute (unrescaled) rank
  difference divided by average network degree.  Each blue curve is a
  network.  The bold black curves represent the mean.}
\label{plot:real_distributions}
\end{figure}

The general forms of these distributions are similar across networks from
different schools.  They also show interesting scaling behavior.  The
widths of the central peaks for both undirected and directed edges, when
measured in terms of raw (unrescaled) rank difference are, to a good
approximation, simply proportional to the average degree of a vertex in the
network.  Figure~\ref{plot:real_distributions} shows these peaks for 78 of
the 84 networks on two plots, for undirected edges (panel~(a)) and directed
edges (panel~(b)), rescaled by average degree, and the approximately
constant width is clear.  (The six networks not shown are all small enough
that the central peaks for the directed edges can be fit by the other
parameters of the model and thus a direct comparison is not appropriate.)
This result indicates that individuals have, roughly speaking, a fixed
probability of being friends with others close to them in rank, regardless
of the size of the community as a whole---as the average number of friends
increases, individuals look proportionately further afield in terms of rank
to find their friends, but are no more likely to be friends with any
particular individual of nearby rank.

\begin{figure}
\begin{center}
\includegraphics[width=6.5cm,clip=true]{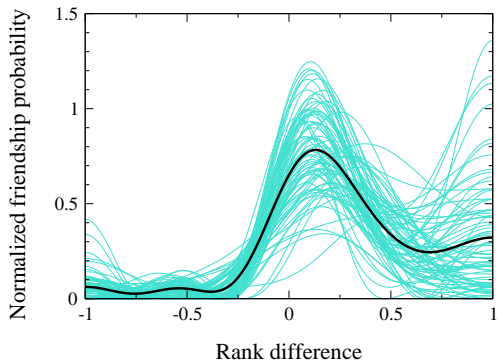}
\end{center}
\caption{The fitted probability function for unreciprocated friendships,
  minus its central peak.  The horizontal axis measures rank difference
  rescaled to run from $-1$ to~1.  Each blue curve is a network.  The bold
  black curve is the mean.}
\label{plot:supposed_distributions}
\end{figure}

Outside of the central peak, i.e.,~for friendships between individuals with
markedly different ranks, there are, to a good approximation, only
unreciprocated friendships, and for these the shape of the probability
distribution appears by contrast to be roughly constant when measured in
terms of the rescaled rank of Fig.~\ref{plot:p_data_show}, which runs from
$-1$ to~1.  This probability, which is equal to the function~$\beta(z)$
with the central Gaussian peak subtracted, is shown in
Fig.~\ref{plot:supposed_distributions} for the same 78 networks, rescaled
vertically by the average probability of an edge to account for differing
network sizes, and again the similarity of the functional form across
networks is apparent, with low probability in the left half of the plot,
indicating few claimed friendships with lower-ranked individuals, and
higher probability on the right.  The roughly constant shape suggests that,
among the unreciprocated friendships, there is, for example, a roughly
constant probability of the lowest-ranked student in the school claiming
friendship with the highest-ranked, relative to other students, no matter
how large the school may be.

The emerging picture of friendship patterns in these networks is one in
which reciprocated friendships appear to fall almost entirely between
individuals of closely similar rank.  A significant fraction of the
unreciprocated ones do the same, and moreover show similar scaling to their
reciprocated counterparts, but the remainder seem to show a quite different
behavior characterized by different scaling and by claims of friendship by
lower-ranked individuals with substantially higher-ranked ones.

\subsection*{Discussion}
\label{Sec:Discussion}
Taking the results of the previous section as a whole, we conjecture that
the rankings discovered by the analysis correlate, at least approximately,
with social status.  If we assume that reciprocated friendships---almost
all of which fall in the central peak---correspond to friendships in the
conventional sense of mutual interaction, then a further conjecture, on the
basis of similar statistics, is that the unreciprocated friendships in the
central peak are also mutual but, for one reason or another, only one side
of the relationship is represented in the data.  One explanation why one
side might be missing is that respondents in the surveys were limited to
listing only five male and five female friends, and so might not have been
able to list all of their friendships.

On the other hand, one might conjecture that the unreciprocated claims of
friendship with higher-ranked individuals, those in the tail of the
distribution in Fig.~\ref{plot:p_data_show}b, correspond to
``aspirational'' friendships, hopes of friendship with higher-ranked
individuals that are, at present at least, not returned.  Note also how the
tail falls off with increasing rank difference: individuals are more likely
to claim friendship with others of only modestly higher rank, not vastly
higher.

\begin{figure}
\begin{center}
\includegraphics[width=\columnwidth]{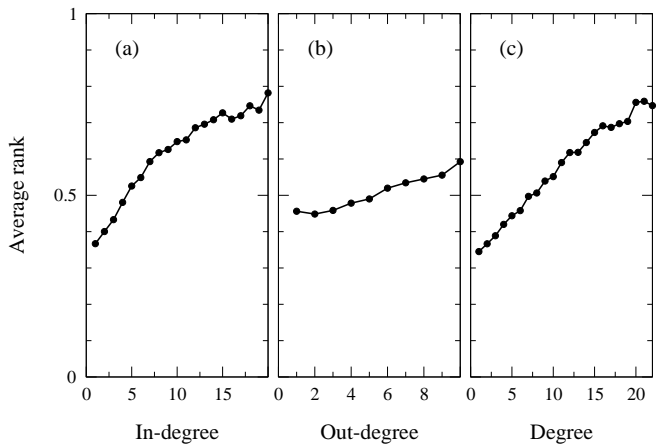}
\end{center}
\caption{Plots of rescaled rank versus degree, averaged over all
  individuals in all networks for (a)~in-degree, (b)~out-degree, and
  (c)~the sum of degrees.  Measurement errors are comparable with or
  smaller than the sizes of the data points and are not shown.}
\label{plot:degree_correlations}
\end{figure}

One way to test these conjectures is to look for correlations between the
rankings and other characteristics of individuals in the networks.  For
instance, it is generally thought that social status is positively
correlated with the number of people who claim you as a
friend~\cite{Hallinan88,Dijkstra10}.
Figure~\ref{plot:degree_correlations}a tests this by plotting average rank
over all individuals in all networks (averaged in the posterior
distribution of Eq.~\eqref{eq:likelihood}) as a function of network
in-degree (the number of others who claim an individual as a friend).  As
the figure shows, there is a strong positive slope to the curve, with the
most popular individuals being nearly twice as highly ranked on average as
the least popular.  Figure~\ref{plot:degree_correlations}b shows the
corresponding plot for out-degree, the number of individuals one claims as
a friend, and here the connection is weaker, as one might expect---claiming
many others as friends does not automatically confer high status upon an
individual---although the correlation is still statistically significant.
Figure~\ref{plot:degree_correlations}c shows rank as a function of total
degree, in-degree plus out-degree, which could be taken as a measure of
total social activity, and here again the correlation is strong.  For all
three panels the correlations are significant, with $p$-values less than
$0.001$.

\begin{figure}
\begin{center}
\includegraphics[width=\columnwidth,clip=true]{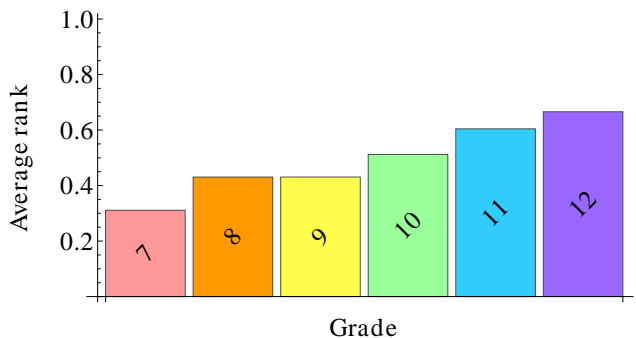}
\end{center}
\caption{Rescaled rank as a function of school grade, averaged over all
  individuals in all schools.}
\label{plot:meta_ranks}
\end{figure}

In addition to the network structure itself, we have additional data about
each of the participants, including their age (school grade), sex, and
ethnicity.  The distributions of rank for each sex and for individual
ethnicities turn out to be close to uniform---a member of either sex or any
ethnic group is, to a good approximation, equally likely to receive any
rank from 1 to~$n$, indicating that there is essentially no effect of sex
or ethnicity on rank.  (A Kolmogorov--Smirnov test does reveal deviations
from uniformity in some cases, but the deviations are small, with KS
statistics $D<0.08$ in all instances.)  Age, however, is a different story.
Figure~\ref{plot:meta_ranks} shows the rescaled rank of individuals in each
grade from 7 to~12, averaged over all individuals in all networks, and here
there is a clear correlation.  Average rank increases by more than a factor
of two from the youngest students to the oldest (a one-way ANOVA gives
$p<0.001$).  Since older students are generally acknowledged to have higher
social status~\cite{Coleman61}, this result lends support to the
identification of rank with status.  A further interesting wrinkle can be
seen in the results for the 8th and 9th grades.  Unlike other pairs of
consecutive grades, these two do not have a statistically significant
difference in average rank (a $t$-test gives $p>0.95$).  This may reflect
the fact that the 8th grade is the most senior grade in the feeder
junior-high schools, before students move up to high school.  When they are
in the 8th grade, students are temporarily the oldest (and therefore
highest status) students in school and hence may have a higher rank than
would be expected were all students in a single school together.

\begin{figure}
\begin{center}
\includegraphics[width=\columnwidth,clip=true]{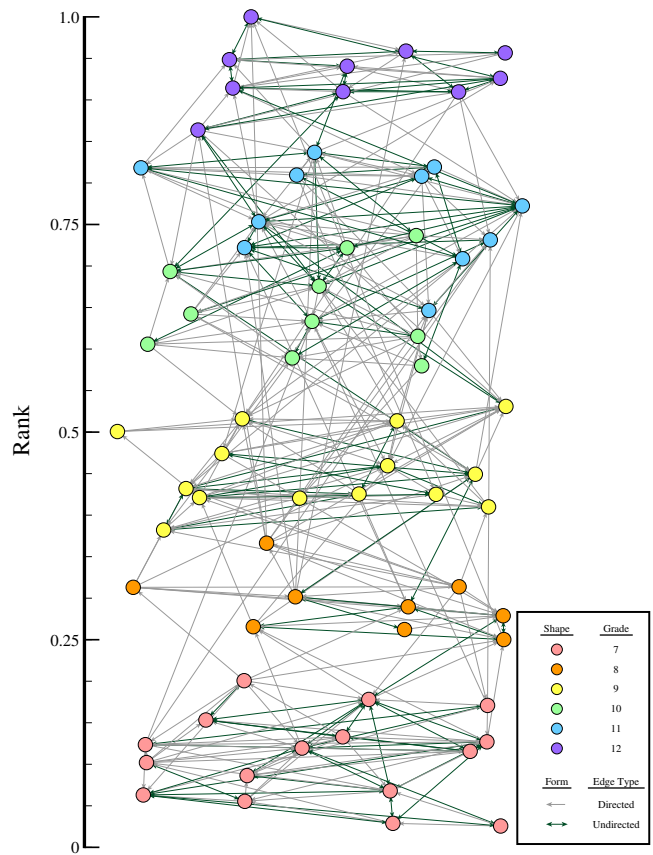}
\end{center}
\caption{A sample network with (rescaled) rank on the vertical axis,
  vertices colored according to grade, and undirected edges colored
  differently from directed edges.  Rank is calculated as an average within
  the Monte Carlo calculation (i.e.,~an average over the posterior
  distribution of the model), rather than merely the maximum-likelihood
  ranking.  Note the clear correlation between rank and grade in the
  network.}
\label{plot:network}
\end{figure}

Finally, in Fig.~\ref{plot:network} we show an actual example of one of the
networks, with nodes arranged vertically on a scale of inferred rank and
colored according to grade.  The increase of rank with grade is clearly
visible, as is the fact that most undirected edges run between individuals
of similar rank (and hence run horizontally in the figure).

\subsection*{Conclusions}
\label{Sec:Conclusion}

In this paper, we have analyzed a large set of networks of friendships
between students in American high and junior-high schools, focusing
particularly on the distinction between friendships claimed by both
participating individuals and friendships claimed by only one individual.
We find that students can be ranked from low to high such that most
unreciprocated friendships consist of a lower-ranked individual claiming
friendship with a higher-ranked one.  We have developed a
maximum-likelihood method for inferring such ranks from complete networks,
taking both reciprocated and unreciprocated friendships into account, and
we find that the rankings so derived correlate significantly with
traditional measures of social status such as age and overall popularity,
suggesting that the rankings may correspond to status.  On the other hand,
rankings seem to be essentially independent on average of other
characteristics of the individuals involved such as sex or ethnicity.

There are a number of questions unanswered by our analysis.  We have only
limited data on the personal characteristics of participants.  It would be
interesting to test for correlation with other characteristics.  Are
rankings correlated, for instance, with academic achievement, number of
siblings or birth order, number of Facebook friends, after-school
activities, personality type, body mass index, wealth, or future career
success?  There is also the question of why a significant number of
apparently close friendships are unreciprocated.  One idea that has
appeared in the literature is that some directed edges may correspond to
new, temporary, or unstable friendships, which are either in the process of
forming and will become reciprocated in the future, or will disappear over
time~\cite{Hallinan88,Sorensen76}.  Evidence suggests that in practice
about a half of the unreciprocated friendships do the former and a half the
latter, and it is possible that the two behaviors correspond to the two
classes of directed edges we identify in our analysis.  A test of this
hypothesis, however, would require time-resolved data---successive
measurements of friendship patterns among the same group of
individuals---data which at present we do not possess.  Finally, there are
potential applications of the statistical methods developed here to other
directed networks in which direction might be correlated with ranking, such
as networks of team or individual competition~\cite{Stefani97,PN05b} or
dominance hierarchies in animal communities~\cite{Drews93,DeVries98}.

\begin{acknowledgments}
  The authors thank Carrie Ferrario, Brian Karrer, Cris Moore, Jason
  Owen-Smith, Bethany Percha, and Claire Whitlinger for useful comments and
  suggestions.  This work was funded in part by the National Science
  Foundation under grant DMS--1107796 and by the James S.  McDonnell
  Foundation.
\end{acknowledgments}

\end{document}